\documentclass[useAMS]{mn2e}

\usepackage{graphicx}

\title[VLBI detection of an Infrared-Faint Radio Source]
{ VLBI detection of an Infrared-Faint Radio Source}

\author[Norris et al.]
	{Ray P. Norris $^{1}$, Steven Tingay$^{2}$, Chris Phillips$^{1}$,     
	Enno Middelberg$^{1}$, Adam Deller$^{2}$, 
\newauthor Philip N. Appleton$^{3}$\\
$^{1}$ CSIRO Australia Telescope National Facility, PO Box 76, Epping, NSW 1710, Australia\\
$^{2}$ Swinburne University of Technology, PO Box 218, Hawthorn, Victoria 3122, Australia\\
$^{3}$ {\it Spitzer} Science Center, California Institute of Technology, 1200 E.California Blvd., Pasadena, CA 91125, USA\\
}

\begin{document}


\pagerange{\pageref{firstpage}--\pageref{lastpage}} \pubyear{2002}

\maketitle

\label{firstpage}

\begin{abstract}

Infrared-Faint Radio Sources represent a new and unexpected class of object which is bright at radio wavelengths but unusually faint at infrared wavelengths. If, like most mJy radio sources, they were either conventional active or star-forming galaxies in the local Universe, we would expect them to be detectable at infrared wavelengths, and so their non-detection by the {\it Spitzer}  Space Telescope  is surprising. Here we report the detection of one of these sources using Very Long Baseline Interferometry, from which we conclude that the sources are driven by Active Galactic Nuclei. We suggest that these sources are either normal radio-loud quasars at high redshift or abnormally obscured radio galaxies. 

\end{abstract}


\section{Introduction}

Over the last two years, in collaboration with the {\it Spitzer}  Wide-area Infrared Extragalactic Survey (SWIRE) team (Lonsdale et al. 2003) and other international groups, we have conducted the Australia Telescope Large Area Survey (ATLAS) of the Chandra Deep Field South (CDFS) and European Large-Area
ISO Survey - S1 (ELAIS-S1) regions, with the aim of producing the widest (7 square degrees) deep ($ \sim 15 \mu$Jy rms) radio survey ever attempted, to help understand the formation and evolution of galaxies in the early Universe. Our survey is quite different in scope from other radio surveys which target a small area to high sensitivity, in that our survey is large enough to uncover rare classes of object, identify cosmological structure, and overcome cosmic variance. 

The key scientific goals are to determine:  

\begin{itemize} 

\item The history of galaxy formation, including the global Star Formation History (SFH),

\item The evolutionary relationship between galaxies and Active Galactic Nuclei (AGN),

\item The changing balance between obscured and unobscured activity over the lifetime of the Universe, 

\item The spatial distribution and clustering of evolved galaxies, starbursts, and AGN, and the evolution of their clustering in the key redshift range from $0.5<z < 3$. 

\end{itemize}

We are now about half-way through the observations, and have completed an analysis of the radio data so far, resulting in images (Afonso et al. 2006; Norris et al. 2006, hereafter N06; Middelberg et al. 2007) with an rms ranging from 14 to 40$\mu$Jy. Our analysis includes source extraction, identification with {\it Spitzer}  and optical data, optical spectroscopy, and fitting spectral energy distributions (SEDs) to produce photometric redshifts and host galaxy classifications. 

Pivotal to our science goals is the ability to distinguish AGN activity from star-formation activity. The radio observations are important because they are unaffected by the heavy dust extinction which is found in the most active galaxies, and are particularly effective at detecting AGN buried within dusty galaxies. For example, some of the ATLAS sources have an SED of starburst galaxies, but a radio luminosity and morphology characteristic of a radio-loud AGN (Norris et al 2007).

An unexpected result has been the discovery of a small class of objects which we call Infrared-Faint Radio Sources, or IFRS. These sources are bright at radio wavelengths but are not detected  in the SWIRE infrared observations. They are also rare: only 60 have been found among the 1600 radio sources identified in the ATLAS survey. A typical IFRS, which happens to be the one detected in this paper, is shown in Figure 1. Stacking of {\it Spitzer}  images at the position of the IFRS has failed to show infrared counterparts, imposing low limits on the IR flux densities, and showing that these are not simply objects which fall just below the {\it Spitzer}  sensitivity limit, but either represent a distinct class of object, or else follow a long-tailed distribution in 3.6 $\mu$m flux density (N06).  

\begin{figure*}
\includegraphics[scale=0.3, angle=0]{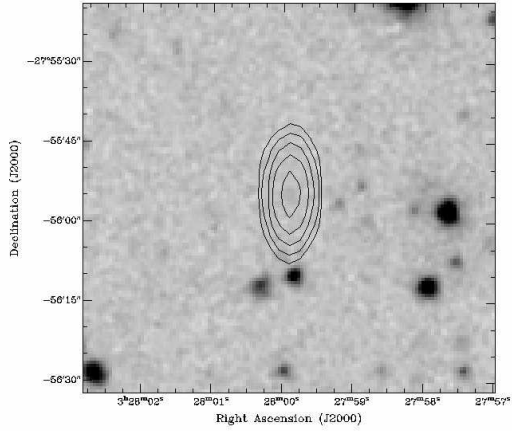}
\caption{20 cm radio image (contours) of the IFRS S114 superimposed on the SWIRE 3.6$\mu$m image (greyscale). Contours are at intervals 0f 0.5, 1, 2, 3, 4, mJy} 
\end{figure*}

Assuming the SED of known classes of galaxy, a 5 mJy radio source in the local Universe should produce a detectable {\it Spitzer}  source, regardless of whether it is generated by star formation or AGN. Suggestions to explain the IFRS have included heavily obscured AGNs, very high-redshift AGNs, and single radio lobes from otherwise undistinguished, and spatially separate, host galaxies. An interesting possibility is that we are catching a starburst or AGN in a transition phase in which electrons have been accelerated to produce radio emission, but there is insufficient hot dust to produce infrared emission. Alternatively, they could be "relic" active galaxies, in which case they will have an extended morphology and a very steep spectral index. Clearly, determining the nature of these objects may be an important step in understanding the early evolution of galaxies. 

A key question is whether the radio emission is generated by an AGN or by star-formation activity. Very Long Baseline Interferometry (VLBI) is a powerful discriminant because AGN's often have compact high-brightness radio cores which are detectable by VLBI, whereas the extended emission characteristic of star-forming galaxies tends to be resolved out and therefore undetectable by VLBI. 

However, we acknowledge that VLBI is not always an unambiguous discriminant because (a) some star –forming regions have radio supernovae of a sufficiently high brightness temperature to be detectable by VLBI, and (b) low-luminosity AGN may be simply too weak to be detectable with current VLBI techniques. In particular, Kewley et al (2000) showed that supernovae in starburst galaxies could masquerade as AGN in VLBI observations. However, the VLBI sources detected here have a higher luminosity than the supernovae sources observed by Kewley et al., and so are unquestionably AGN cores.

\section{Observations and Data Reduction}

The VLBI observations took place over a period of twelve hours on 14 March 2006. Four radio telescopes were used for this experiment: the 64-m antenna of the Australia Telescope National Facility (ATNF)
near Parkes; a phased array at the ATNF Australia Telescope Compact Array (ATCA)
near Narrabri; the ATNF Mopra 22-m antenna near Coonabarabran;
and the University of Tasmania's 26-m antenna near Hobart.  The Narrabri phased array consisted of five antennas of the Australia Telescope Compact Array configured in an ultra compact (EW367) configuration. We observed a small sample of sources, shown in Table 1.  Source names and characteristics are taken from N06.

\begin{table*}
 \centering
  \caption{Sources observed.}
  \begin{tabular}{lllll}
  \hline
Short & Full & classi-  & peak ATCA flux & Total detected VLBI \\
Name  & name & fication & density (mJy/beam)  & flux density (mJy)\\
 \hline
 
S194	& ATCDFS J032928.59-283618.8	& IFRS 	& 4.6 	& $<$ 1.0	\\
S114	& ATCDFS J032759.89-275554.7	& IFRS 	& 5.7 	& 5.0	\\
S201	& ATCDFS J032933.82-284140.2	& SF galaxy & 11.3	& $<$ 2.0 	\\
S091	& ATCDFS J032737.74-280130.0	& AGN1	& 36.2	& 17.6 	\\
S437	& ATCDFS J033226.97-274106.7	& AGN1	& 13.1	& $<$ 2.0 \\
S089	& ATCDFS J032734.01-284621.3	& Calibrator & 131.2	& 126.0 \\
S145	& ATCDFS J032836.58-284145.5	& Calibrator & 798.4	& $\sim$20.0 \\


\hline
\end{tabular}
\end{table*}

Four 16-MHz  bands, two each from right and left circular polarisation, were recorded 
to disk. The data were two-bit sampled, giving a total data rate of 256 
Mbps per station.  The central observing frequency was 1.650 GHz. We observed in a forty-minute cycle, switching between the target sources, including the phase-reference calibrator sources, every twenty minutes. 

The data were correlated with the Swinburne software correlator (Deller et al. 2007), initial calibration performed using measured system parameters at each telescope, and the data then imported to AIPS for final calibration and data reduction.

Strong fringes were detected on all baselines on the calibrator sources and two target sources. 

Because we had no {\it a priori} information that any of the sources were unresolved on these baselines, the only calibration applied to the data was to use the measured sensitivity for each antenna, to convert (using AIPS) the measured visibilities into milliJansky of correlated flux density on each baseline. The data were then edited to remove sections of data for which the antennas were not on source.

Where possible, the data were then self-calibrated and imaged using DIFMAP (Shepherd, 1997). However, some sources were too weak on some baselines to image. We found that the most sensitive way of detecting and measuring correlated flux density was to perform a fringe-rate transform of the u-v series data using the AIPS task FRPLT. A typical output for one scan is shown in Figure 2. In this way, the correlated flux density was measured for each source in each scan, and the resulting flux density as a function of u-v distance (or antenna separation) is shown for three sources in Figure 3.

\begin{figure*}
\includegraphics[scale=0.5, angle=0]{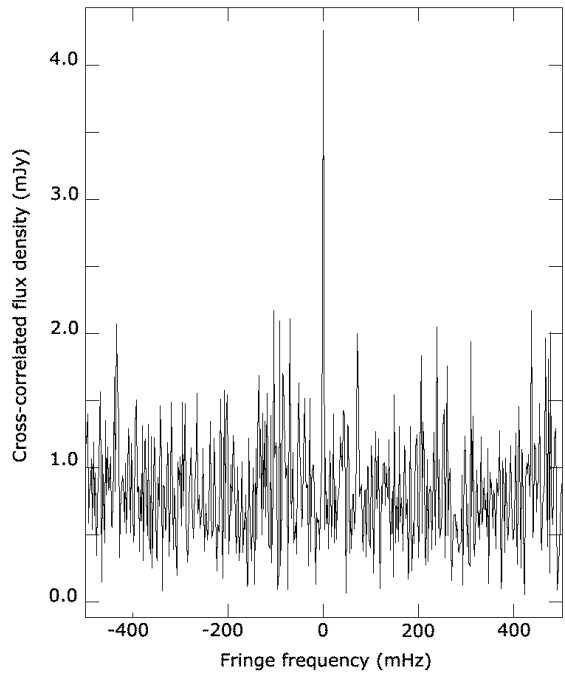}
\caption{A fringe rate spectrum on the Parkes-Narrabri baseline of a typical scan of 7 minutes of data on the IFRS source S114.} 
\end{figure*}

Because the only calibration that has been applied to these data is the antenna-based sensitivity, the amplitudes could in principle be significantly in error. The consistency of the measured correlated flux densities  across the three sources and across all baselines used in the experiment gives us confidence that this is not the case.

\begin{figure*} 
\includegraphics[scale=0.6, angle=0]{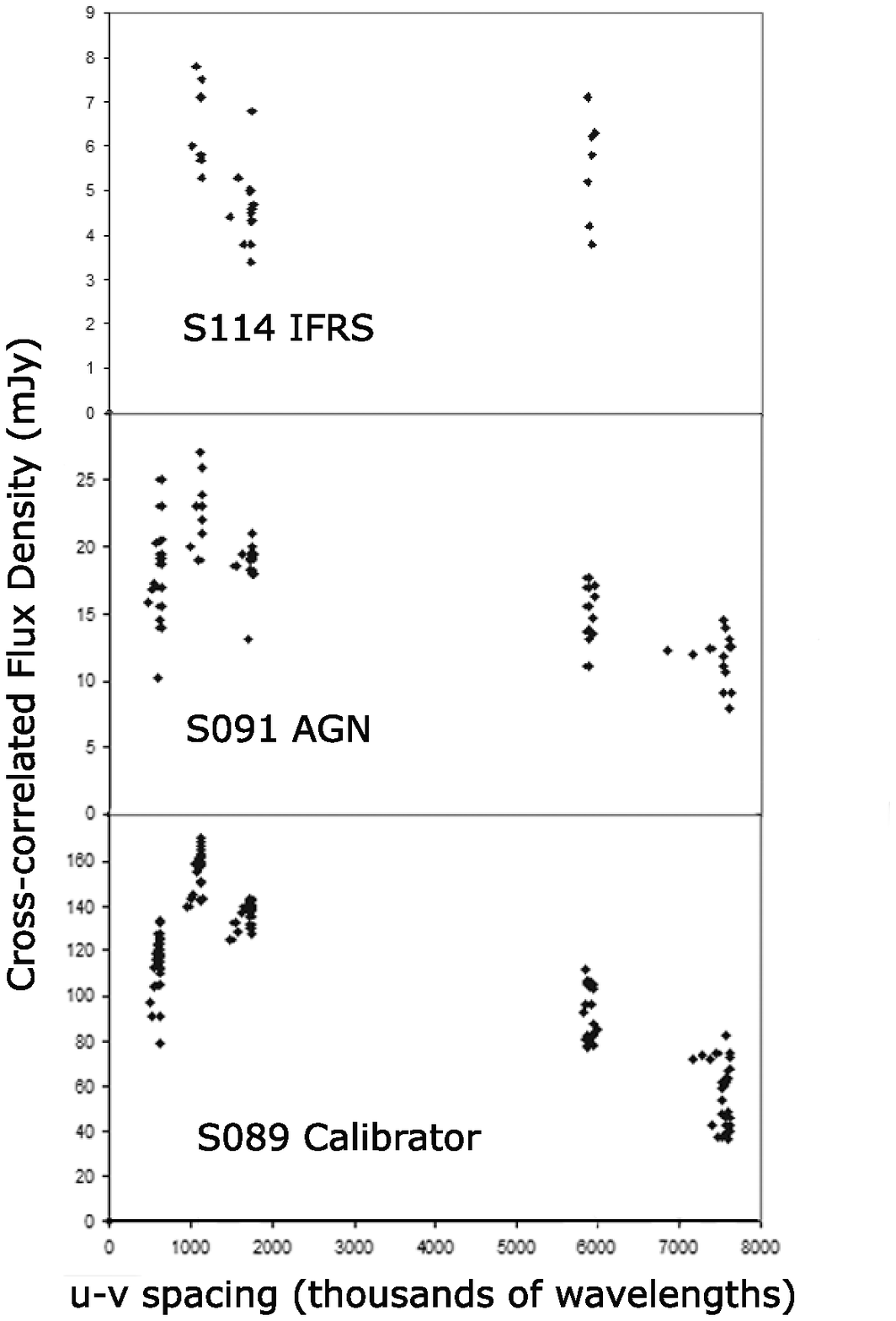}
\caption{Plots of correlated radio flux density as a function of uv-spacing for three of the sources discussed in the text. }
\end{figure*}

\section{Results}

Table 1 lists the observed sources, including their 
classifications and measured flux densities on ATCA and VLBI scales. 

\subsection{The undetected and confused sources}
We failed to detect any emission from S194 (an IFRS), S201 (a star-forming galaxy, which was included as a control), and S437 (a QSO). Non-detection on these objects does not necessarily imply that there is no radio core, but merely tells us that any compact AGN core is below our sensitivity limit, and that the radio emission detected from these sources in the ATLAS survey is dominated by emission from an extended region. 

The observed flux density of the strongest source in our field, S145, which we included as a calibrator, varied rapidly from scan to scan, indicating that the radio emission is dominated by much larger-scale structure. The erratic nature of these fluctuations suggests a complexity which is beyond the scope of the available data, and we were unable to produce a reliable image using DIFMAP. We do not therefore consider this source any further.

\subsection{The AGN S089}
The other calibrator in our observations, S089, had a detected amplitude which varied only slightly from scan to scan, implying that this source is dominated by a compact AGN core, although a reduced visibility at long baselines suggests that it is partially resolved. 

\begin{figure*} 
\includegraphics[scale=0.4, angle=0] {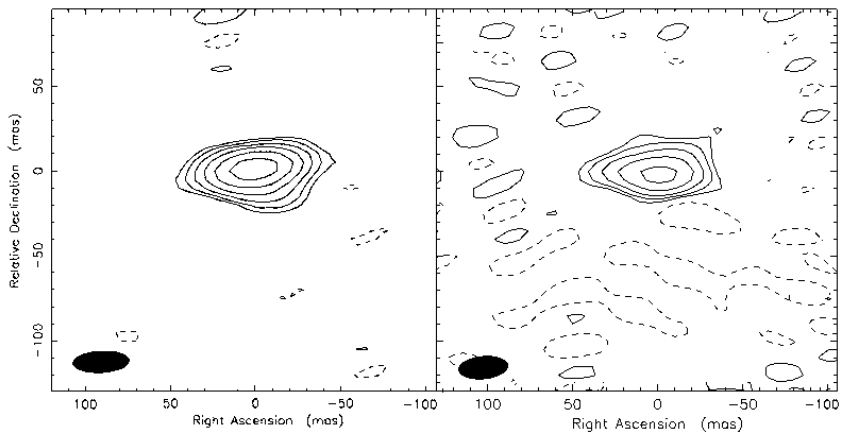}
\caption{(Left)VLBI image of the AGN S089. Contours are at -2, 2, 4, 8, 16, 32, 64 \% of the peak flux of 91 mJy/beam. (Right) VLBI image of the AGN S091. Contours are at -5, 5, 10, 20, 40, 80 \% of the peak flux of 11.7 mJy/beam. In both cases, the restoring beam as shown as a dark ellipse in the lower left-hand corner.}
\end{figure*}

S089 is a strong (133 mJy) AGN identified during the ATLAS program (N06) and for which we have not yet measured a redshift. Strong fringes were obtained on all baselines in the VLBI experiment, enabling us to self-calibrate (selfcal) and image it using DIFMAP, resulting in the image shown in Figure 4 (left). It has a total flux integrated over the source of 126 mJy, and a peak flux density of 91 mJy/beam. The total VLBI flux density is close to the values measured on the ATCA by N06 of the total flux density (133 mJy) and the peak flux density (131 mJy/beam), suggesting that this source has no significant extended emission beyond that seen in this image, making it an excellent calibrator for future VLBI experiments. We attribute the slight increase in detected VLBI flux density above 126 mJy in Figure 3 at short baselines to the calibration uncertainty of our VLBI data, which were removed by selfcal to produce the image in Fig 4. 

\subsection{The AGN S091}
The source S091 was classified by N06 as an AGN on the basis of its radio-FIR ratio, and was also tentatively identified as a quasar on the basis of spectroscopy, which placed it at a redshift of 1.174. The ATCA image published by N06 clearly shows that the source is elongated, at a position angle of 54 degrees, and its peak flux density of 36 mJy/beam is significantly lower than its integrated flux density of 96 mJy. 

Strong fringes were obtained on all baselines in the VLBI experiment, enabling us to selfcal and image it using DIFMAP, resulting in the image shown in Figure 4 (right). It has a peak flux density measured from the image of 11.7 mJy/beam and a total flux of 17.6 mJy. This VLBI flux density is about half the ATCA peak flux density measured by N06, and its VLBI core is only slightly resolved on the longest VLBI baselines. All these observations are consistent with S091 being a classical FRI or FRII radio-loud quasar, with an 18mJy compact core at its nucleus and extended jets and lobes accounting for the remaining 78 mJy. The total radio luminosity of the source is $8 \times 10^{26} W Hz^{-1}$, which places it in the FRII luminosity regime.

\subsection{The IFRS S114}
The IFRS source S114 was clearly detected on several baselines at all hour angles, as shown in Figures 2 and 3, but it is not sufficiently strong on the least sensitive baselines to be able to derive an image from the data. We therefore draw our results and conclusions on this source entirely from the amplitudes of the visibility data. We note that the consistency of measurements on the two sources S089 and S091 suggest that the calibration of the visibility amplitudes is correct to about 25\%.

S114 has a measured VLBI flux density of about 5 mJy, which is comparable to the peak ATCA flux density of 5.7 mJy/be, and the total ATCA flux density of 7.2 mJy, measured by N06, who found it marginally resolved. The source appears to keep the same correlated flux density even on the longest baselines, implying that it has a core size of less than 0.03 arcsec.

\begin{table*}
 \centering
  \caption{Expected flux densities from 3C273 and Cygnus A if they were moved to z=1 and z=7, including k-corrections }
  \begin{tabular}{llll}
  \hline
Source      & Redshift & 20cm Flux & 3.6$\mu$m flux \\
            &         & (mJy)     & ($\mu$Jy)  \\
 \hline
 
3C273	      & 1       & 713 	    & 635 	\\
3C273  	& 7       & 4.5 	    & 2.8 \\
Cygnus-A	& 1       & 695       & 22 	\\
Cygnus-A	& 7	    & 1.9       & 0.04 	\\

\hline
\end{tabular}
\end{table*}

\section{Discussion}

The radio core in the S114 IFRS appears unresolved on VLBI baselines, implying an angular core size of less than 0.03 arcsec, corresponding to 260 pc or less at any redshift. If the marginal detection of extended structure reported by N06 is correct, then S114 consists of an unresolved 5 mJy core with a size of less than 0.03 arcsec, and a brightness temperature  $> 6 (1+z) \times 10^{6} K$, surrounded by a low-brightness structure with a flux density of about 2 mJy and a size of about 4.6 arcsec. This latter structure might well consist of extended lobes or jets ejected from the compact core. 

As the redshift of this object is unknown, it is difficult to compare it with known galaxies. We therefore consider its properties at two possible redshifts.

If the source is at z=1, its lobes have a linear extent of 37 kpc, and it has a luminosity of 4$\times 10^{25} W Hz^{-1}$. These characteristics place it well within the normal span of radio galaxies, from which it differs only in that it is not detected in the SWIRE catalog.

If the source is at z=7, the lobes have a linear extent of 25 kpc, and the source has a total luminosity of $4 \times 10^{27} W Hz^{-1}$. Assuming a spectral index of -0.7, and including k-corrections, this is equivalent to a rest-frame 5 GHz luminosity of $7 \times 10^{27} W Hz^{-1}$. This luminosity lies at the high end for known radio galaxies (e.g. O`Dowd et al, 2002) but is not abnormal. Non-detection of a radio galaxy at z=7 would not be unusual for SWIRE, as will be shown below. In this case, the only abnormality is that it lies at z=7.

Table 2 shows the effect of moving a 3C273-like source and a Cygnus-A source to z=1 and z=7. The values shown in the table include k-corrections and are based on a standard cosmology with H$_{0}$=71 km s$^{-1}$ Mpc$^{-1}$, $\Omega_{m}$=0.27, $\Omega_{vac}$=0.73.  Given that the quoted SWIRE detection sensitivity at 3.6$\mu$m is about 5 $\mu$Jy, either source at z=1 would be too strong at both wavelengths to reproduce S114, but 3C273 at z=7 would observationally closely resemble S114. A straightforward explanation of the IFRS class is therefore simply that they are radio-loud quasars or galaxies at high redshift.

Alternatively, an IFRS could be produced by reducing the power of a 3C273-type AGN, and placing it at a lower redshift. However, the SWIRE survey sensitivity can detect normal L-star galaxies at redshifts up to z $\sim$ 1, and so it is unlikely that the host galaxy of even a low-power AGN at low redshift would be undetected by SWIRE.

We also cannot rule out the possibility that the IFRS's consist of lower-redshift, lower-luminosity galaxies in which the infrared emission is shielded by many magnitudes of dust extinction extending even to the mid-infrared. Other models involving transitional stages, or combinations of these factors are also possible. Unfortunately, the data presented here do not readily distinguish between these models. We can however be confident that S114 contains an AGN, and that it is consistent with a normal radio-loud quasar at high redshift.

\section{Conclusion}

We have detected the IFRS S114 with VLBI, from which we conclude that the radio emission is generated by an Active Galactic Nucleus. The absence of a detection of the host galaxy by {\it Spitzer}  may be interpreted as implying that either (a) the source is a normal radio-loud quasar or radio galaxy at high redshift, or (b) the source is abnormally obscured by dust. We cannot rule out the more exotic possibility that the source represents a new class of galaxy, such as a high-luminosity AGN in a low-luminosity host galaxy, or one in a stage of transition, but consider this less likely.

\section*{Acknowledgments}

We thank the Australian LBA team whose continuing efforts make observations like this possible, and especially the University of Tasmania for making the Hobart antenna available. Like countless other papers, this one has  made extensive use of Ned Wright's Cosmology Calculator (http://www.astro.ucla.edu/~wright/CosmoCalc.html ) and we thank and applaud Ned Wright for making such a valuable tool available to the community.

The Australia Telescope
is funded by the Commonwealth of
Australia for operation as a National Facility managed by CSIRO. This research has made use of
the NASA/IPAC Extragalactic Database (NED) which is operated by the Jet Propulsion Laboratory,
California Institute of Technology, under contract with the National Aeronautics and Space
Administration. ATD 
is supported by a Swinburne University Chancellor's Research scholarship 
and a CSIRO postgraduate scholarship

\label{lastpage}


\begin{thebibliography}{99}
\bibitem[Afonso et al.(2006)]{2006AJ....131.1216A} Afonso, J., Mobasher, 
B., Koekemoer, A., Norris, R.~P., \& Cram, L.\ 2006, AJ, 131, 1216 

\bibitem[Deller et al.(2007)]{2007astro.ph..2141D} Deller, A.~T., Tingay, 
S.~J., Bailes, M., \& West, C.\ 2007, ArXiv Astrophysics e-prints, 
arXiv:astro-ph/0702141 

\bibitem[Kewley et al.(2000)]{2000ApJ...530..704K} Kewley, L.~J., Heisler, 
C.~A., Dopita, M.~A., Sutherland, R., Norris, R.~P., Reynolds, J., \& 
Lumsden, S.\ 2000, ApJ, 530, 704 

\bibitem[Lonsdale et al.(2003)]{2003PASP..115..897L} Lonsdale, C.~J., et 
al.\ 2003, PASP, 115, 897

\bibitem []{} Middelberg, E., et al., 2007, submitted to AJ

\bibitem[Norris et al.(2006)]{2006AJ....132.2409N} Norris, R.~P., et al.\ 
2006, AJ, 132, 2409 (N06)

\bibitem[Norris et al.(2007)]{2007astro.ph..1360N} Norris, R.~P., 
Middelberg, E., \& Boyle, B.~J.\ 2007, ArXiv Astrophysics e-prints, 
arXiv:astro-ph/0701360

\bibitem[O'Dowd et al.(2002)]{2002ApJ...580...96O} O'Dowd, M., Urry, C.~M., 
\& Scarpa, R.\ 2002, ApJ, 580, 96 

\bibitem[Shepherd(1997)]{1997ASPC..125...77S} Shepherd, M.~C.\ 1997, ASP 
Conf.~Ser.~125: Astronomical Data Analysis Software and Systems VI, 125, 77

\end{thebibliography}
\end{document}